\documentclass[conference, a4paper]{IEEEtran}
\IEEEoverridecommandlockouts
\usepackage{cite}
\usepackage{amsmath,amssymb,amsfonts}
\usepackage{booktabs}
\usepackage{algorithmic}
\usepackage{multirow}
\usepackage{graphicx}
\usepackage{subfig}
\usepackage{textcomp}
\usepackage{xcolor}
\usepackage{makecell}
\def\BibTeX{{\rm B\kern-.05em{\sc i\kern-.025em b}\kern-.08em
    T\kern-.1667em\lower.7ex\hbox{E}\kern-.125emX}}

\begin{document}

\title{VI-SLAM2tag: Low-Effort Labeled Dataset Collection for Fingerprinting-Based Indoor Localization\\
\thanks{The dataset can be found at DOI:10.5281/zenodo.6801310. \\The implementations are hosted at https://github.com/laskama.}
}

\author{\IEEEauthorblockN{Marius Laska, Till Schulz, Jan Grottke, Christoph Blut, J\"org Blankenbach}
\IEEEauthorblockA{\textit{Geodetic Institute and Chair for Computing in Civil Engineering \& Geo Information Systems} \\
\textit{RWTH Aachen University}\\
Aachen, Germany \\
marius.laska@gia.rwth-aachen.de}
}

\maketitle

\begin{abstract}
Fingerprinting-based approaches are particularly suitable for deploying indoor positioning systems for pedestrians with minimal infrastructure costs. The accuracy of the method, however, strongly depends on the quality of collected labeled fingerprints within the calibration phase, which is a tedious process when done manually in a static fashion. We present \mbox{VI-SLAM2tag}, a system for auto-labeling of dynamically collected fingerprints using the visual-inertial simultaneous localization and mapping (VI-SLAM) module of ARCore.
ARCore occasionally updates its internal coordinate system. Mapping the entire trajectory to a target coordinate system via a single transformation thus results in large drift effects.
To solve this, we propose a strategy for determining locally optimal sub-trajectory transformations.
Our system is evaluated with respect to the accuracy of the generated position labels using a geodetic tracking system.
We achieve an average labeling error of roughly 50 cm for trajectories of up to 15 minutes, which is sufficient for fingerprinting-based localization.
We demonstrate this by collecting a multi-floor dataset including WLAN and IMU data and show how it can be used to train neural network based models that achieve a positioning accuracy of roughly 2 m. \mbox{VI-SLAM2tag} and the dataset are made publicly available.
\end{abstract}


\section{Introduction}
Obtaining the position of a person or entity is the key requirement to offer location based services (LBS) such as navigation or point-of-interest (PoI) queries. LBS have become omni-present in the last decade and are integrated into many mobile applications. LBS also have manifold potential applications for indoor environments, such as shopping malls, airports or hospitals \cite{BLM+17}. Presently, for retrieving the location of a person or entity indoors, no gold standard such as global navigation satellite systems for outdoors exists.
The field of indoor localization has been growing for the last few years resulting in various techniques and technologies that differ mainly in their trade-off between accuracy and costs \cite{ZGL19}. Choosing the right positioning technology should be application driven by estimating the specific requirements in terms of required accuracy of the use case, deployability and accompanied costs. Especially, for mass-market and large-scale pedestrian indoor localization, systems are required that provide medium to coarse-grained localization accuracy at minimal deployment effort and infrastructure costs.
One technique that is suitable for such scenarios is fingerprinting. The idea is that a unique fingerprint can be obtained for each location within the building, where the fingerprint consists of multiple signals observed by a sensor of the device which should be localized. The mapping between fingerprints and locations can be predicted by an algorithm. Recently, machine learning (ML) has been extensively used for this purpose \cite{BTP21}.
Training a supervised ML algorithm requires fingerprints labeled with the corresponding position. The accuracy of trained algorithms mainly depends on the quality and quantity of labeled data. When collecting these data manually at fixed reference positions, this can become a tedious task, which increases the deployment effort of such systems.
Several strategies have been proposed to tackle this issue: 1) given a set of manually labeled data, new fingerprints can be constructed (e.g. via a generative model or data augmentation techniques) \cite{AK21a, NCC+21}; 2) by semi-supervised learning a set of manually labeled data can be used in combination with real fingerprints that are not labeled \cite{CA19}; 3) crowdsourcing allows for splitting the labeling effort across multiple participants \cite{JML+16} and 4) auto-labeling can decrease the amount of human intervention that is required \cite{TS21}. Combinations of all strategies are possible. 1) and 2) require a certain amount of labeled data and 3) only distributes the effort by acquiring more manual workforce. Therefore, 4) has the highest potential of lowering the barrier of practical deployments.

Manual labeling can be avoided by automatically obtaining ground truth data during data collection.
Ideally, the ground truth should originate solely from the device (mostly smartphone) which is used to collect the fingerprints. This requires the fusion of several sensors. For example, Zee \cite{RCP+12} utilizes the inertial measurement unit (IMU) of the smartphone and applies pedestrian dead reckoning within a particle filter. An additional sensor, which can drastically improve the localization quality is the smartphone camera. Augmented Reality (AR) frameworks like ARCore (Google) or ARKit (Apple) provide great potential for location estimation, since they offer out-of-the-box visual inertial simultaneous localization and mapping (VI-SLAM), which are optimized for mobile hardware (smartphones).
Still, ARCore cannot be directly used to obtain absolute position tags as it only tracks the position within its local coordinate system. Furthermore, the local coordinate system is occasionally updated to keep consistency, which makes estimating a global transformation for mapping the entire trajectory challenging \cite{Google}.

We present \emph{VI-SLAM2tag}, a system consisting of a smartphone application that leverages ARCore for relative pose tracking and a post-processing approach based on landmark mapping to obtain position tags within a defined target coordinate system.
To handle coordinate system updates of ARCore, we propose a mapping strategy that determines locally optimal sub-trajectory transformations.
Using our system we collect a multi-floor dataset in our university building. The dataset is suitable for training models that require continuously labeled data such as RSS fingerprinting models or learned models for relative position estimation using IMU data \cite{CLW+21}.

The rest of the paper is structured as follows. We begin by discussing related work with respect to reducing the manual labeling effort of fingerprinting-based indoor localization and SLAM-based localization in section \ref{sec:related_work}. Afterward, we describe the VI-SLAM2tag workflow in section \ref{sec:workflow} and introduce the smartphone application by covering the fundamental concepts of ARCore in section \ref{sec:smartphone_app}. In section \ref{sec:postprocessing} we introduce two post-processing strategies for obtaining position tags within the target coordinate system. Our VI-SLAM2tag system is evaluated in section \ref{sec:evaluation} with respect to the accuracy of the obtained position tags.
Finally, in section \ref{sec:dataset} we introduce the dataset that we collected and show the baseline performance of two algorithms. The results are concluded in section \ref{sec:conclusion}.
\begin{figure*}[t!]
    \centering
    \includegraphics[trim=0 320 10 0, clip, width=\linewidth]{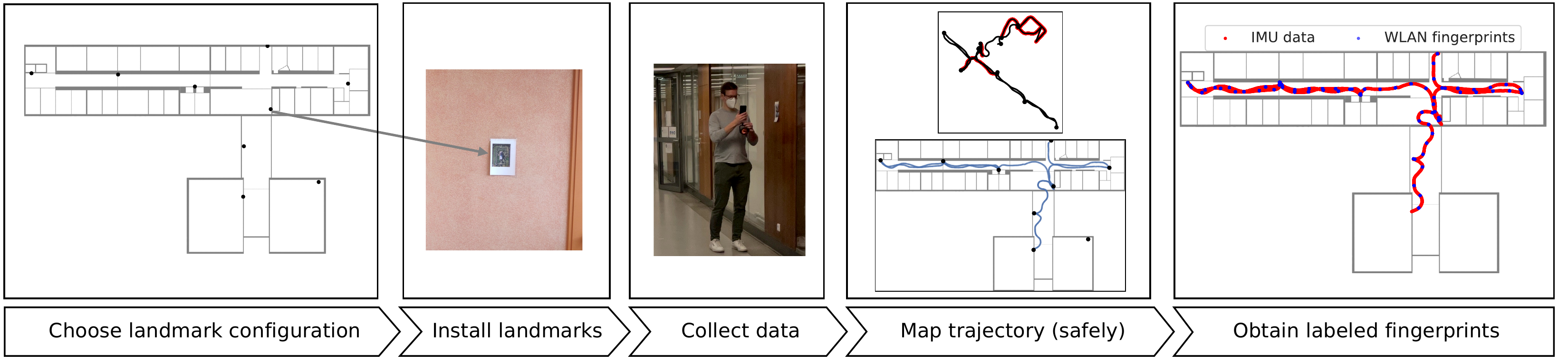}
    \caption{VI-SLAM2tag workflow: 1) Landmark configuration is chosen for a given building floor plan; 2) Landmarks are installed at the chosen locations within the building and positions are registered in BCS; 3) Data is collected while freely walking and scanning passed landmarks; 4) Logged position data in ACS is safely mapped to BCS while detecting failures of the VI-SLAM algorithm; 5) Non-corrupted position data is used to annotate collected data via time-based matching.}
    \label{fig:pipeline}
\end{figure*}

\section{Related work}\label{sec:related_work}

\subsection{Decreasing manual labeling effort for fingerprinting}
While fingerprinting-based indoor localization does not require a dedicated positioning infrastructure, it relies on a rich set of labeled fingerprints to achieve a satisfactory localization accuracy.
The most basic approach for setting up such a fingerprinting database is to collect fingerprints at predetermined reference points in the calibration phase. Using a collection device (smartphone) the surveyor statically stands over the reference position and collects one or more fingerprints.
This procedure is very time-consuming, which makes fingerprinting-based positioning systems tedious to deploy, especially within large-scale environments such as shopping malls.
A large body of research has been focused on reducing the manual labeling effort required for data collection, where works can be mainly grouped into 4 categories:
\subsubsection{Data augmentation/generative models} Given an initial set of labeled fingerprints, the objective can be to artificially enlarge it. Inspired from data augmentation in image classification, one can slightly alter existing fingerprints. Sinha et al. \cite{SLR+19} add noise to certain AP-RSS entries while being compliant to observed statistical quantities. Anagnostopoulos and Kalousis \cite{AK21a} propose proximity-based augmentation. They enlarge the dataset by combining fingerprints that have been collected within close proximity.
Another approach is to train generative models that predict fingerprints even for locations that have not yet been visited before. Njima et al. \cite{NCC+21} train a generative adversarial network which generates fake fingerprints (RSS only) data based on a small set of real collected labeled data. Subsequently, pseudo-labels of the generated RSS fingerprints are predicted.
\subsubsection{Semi-supervised learning} Unlabeled data is much easier to collect, since it does not require human intervention and can be collected while freely walking around the site. Together with a core-set of labeled data it can be used to improve the learning process of neural networks. Chidlovskii and Antsfeld \cite{CA19} train a deep variational autoencoder to utilize a large set of unlabeled data to enhance the performance of various localization models.
\subsubsection{Crowdsourcing} Instead of relying on a single person that collects the labeled data, the surveying process can be split among several participants, possibly non-trained experts. While this drastically reduces the survey time, it introduces further risks of faulty data and requires incentives for user participation \cite{JML+16, JH17, PTU+19}.
\subsubsection{(Semi-)automatic labeling} The labeling effort can be drastically decreased by semi-automatically inferring position tags during site survey. By utilizing additional sensor sources, the location of the sensing device can be determined during data collection. Rai et al. \cite{RCP+12} use a particle filter which utilizes the IMU and achieves further convergence by the already gained knowledge about annotated WLAN fingerprints of already visited locations. As an alternative sensor besides IMU, the smartphone camera offers great potential for relative position estimation via visual odometry. By detecting feature points and tracking them across successive camera frames, the change in position can be inferred. Tomažič and Škrjanc \cite{TS21} perform online calibration of BLE signal strength mapping via Android ARCore.

The first two approaches still require a certain amount of labeled data and crowdsourcing only lowers the effort by acquiring more manual workforce. Therefore, the last category (semi-automatic labeling) has the highest potential of lowering the barrier of practical deployments and can be combined with the other three strategies.

\subsection{VI-SLAM}
Estimating and tracking the pose of a moving object from camera images are widely studied research subjects, also in the field of indoor localization. Prominent application scenarios are autonomous driving systems and robot navigation \cite{CER+21}. The applied methods can be differentiated between visual odometry (VO) and visual simultaneous localization and mapping (V-SLAM). VO derives poses relative to an arbitrary starting pose by comparing the camera movement between successive camera images. V-SLAM is based on VO and extends it by incorporating a 3D map into the process, which is constructed during localization. The map enables optimizing the total trajectory and current pose via loop closure, when revisiting a previous location. For V-SLAM, typically a monocular or stereo camera setup are used. While depth (scale) can be obtained directly using a stereo setup, a monocular setup does not provide this information, so that additional sensors, such as an IMU, must be integrated via sensor fusion. The process is referred to as visual inertial SLAM then (VI-SLAM). While there exist multiple V-SLAM solutions, there are only a few that take inertial data into account. ORB-SLAM \cite{MMT15, CER+21} is a well-known framework for V-SLAM and recently was extended to VI-SLAM, but is not optimized for mobile applications. On smartphones, Augmented Reality (AR) has gained popularity the past years, so that Google and Apple provide the frameworks ARCore \cite{Google} for Android and ARKit \cite{Inc} for iOS to create AR experiences. Both frameworks contain highly optimized VI-SLAM-based pose tracking solutions for mobile devices. Given that fingerprints are mostly collected with smartphones, those frameworks offer great potential for automatically obtaining position tags. However, since they are mostly closed source and not originally intended for localization, it is challenging to design such a system.
It has been shown that computing a mapping by simply aligning the origin and axes of ARCore's local coordinate system and those of the real-world coordinate system does not suffice, since drift of up to 17m per 120m trajectory occurs \cite{FPS+20}.

\section{VI-SLAM2tag workflow}\label{sec:workflow}
The goal of VI-SLAM2tag is to obtain sensor data (WiFi fingerprints and IMU data) that are automatically labeled with the position of collection within a specified \textit{building coordinate system (BCS)}. ARCore is utilized for providing the basis for annotating the sensor data, however, it tracks the smartphone position within its own local coordinate system, which we refer to as \textit{ARCore coordinate system (ACS)} from now on. Simple transformations from ACS to BCS as described in \cite{FPS+20} do cause severe drift effects.
Instead, we propose to estimate a transformation by using corresponding point pairs of both coordinate systems.
This is enabled by detecting manually placed landmarks within the environment of which we exactly know the position within the BCS.
The workflow of VI-SLAM2tag is presented in Figure \ref{fig:pipeline} and is described in the following.

Initially, a landmark configuration has to be chosen for a given floor of the building. Landmarks should be placed after each critical turn a data collector might take when walking around the building. Next, the landmarks need to be installed at the physical locations within the building and their positions within the BCS have to be exactly determined by geodetic surveying (e.g. using a total station). Now, an arbitrary amount of data can be collected while walking around with a smartphone that logs its position in the ACS along with sensor data (e.g. WLAN fingerprints and IMU data) for subsequent annotation. Once the collector passes a landmark, it is scanned via the smartphone to obtain its coordinates in the ACS.
Afterwards, the estimated transformation of the coordinate pairs is used to transform the entire trajectory from ACS to BCS. Finally, the collected sensor data are annotated with the transformed coordinates to obtain a labeled fingerprinting dataset.
In the following we will introduce the design of the smartphone application for data collection in section \ref{sec:smartphone_app} and subsequently explain how we obtain a mapping to the BCS, which is robust against failure of the SLAM-based positioning in section \ref{sec:postprocessing}.

\section{VI-SLAM based data collection (via ARCore)}\label{sec:smartphone_app}

\subsection{Fundamental concepts of ARCore}
In the previous section we identified that ARCore is a suitable solution for designing an easy to use labeled data collection system. In this section we will not cover the exact VI-SLAM algorithm behind ARCore, since its mostly closed source. For a more thorough explanation the reader is referred to the corresponding patent \cite{NLZ17}. Instead, we want to point out the main principles and assumptions of ARCore which make it suitable for the task at hand. Those can be found in the official API documentation \cite{Google}.

ARCore continuously tracks the smartphone's pose within the virtual world and updates its understanding of the virtual world based on the collected data. It uses visual odometry for computing relative pose changes by identifying matching feature points in consecutive image frames and tracking them over time. Since the smartphone camera is monocular, the scale cannot be inferred directly, but is obtained by fusing data of the smartphone's IMU.
In the following, italic words represent concrete objects/concepts within the ARCore framework.
A location within the ACS is represented by a \textit{Pose}. The current position can therefore be obtained by accessing the \textit{Pose} of the \textit{Camera} object. However, as ARCore's understanding of the environment changes over time, it adjusts its model of the world to keep things consistent. This means that the ACS is updated, such that the camera pose might appear at notably different coordinates after the update. Therefore, it is problematic to use the \textit{Pose} object of the \textit{Camera} to obtain the user's relative position across different camera frames, which is the motivation for our proposed mapping strategy introduced in section \ref{sec:local_mapping}.

ARCore introduces a concept called \textit{Anchors} to attach objects to fixed locations within the virtual world. This is mainly done for supporting AR applications. One can place an object on a table and it will appear at the same exact position, even if ARCore updates its ACS. In our approach anchors will be used to represent the fixed locations of the landmarks within the virtual world.

ARCore offers a concept called \textit{Augmented Images} for detecting previously defined 2D images within the virtual world. If such an image is detected it is automatically tracked from now on via placing an anchor at its estimated position. When developing our application, we utilize this concept for detecting our landmarks. Once a landmark appears close within the current camera frame, ARCore will identify it and automatically track it even if it leaves the camera frame subsequently. This allows us to log the position of each detected landmark (after it has been recognized the first time) for each camera pose obtained by the system.

\subsection{VI-SLAM2tag smartphone application}

To implement our data collection application, we extend the sample project \textit{computer\_vision} of the official sample application provided by the arcore-android-sdk \cite{22}. The main activity implements the \textit{GLSurfaceView.Renderer} interface, which consists of the \textit{onDrawFrame} function that is called once a new camera frame is available. Via the \textit{Session} object of ARCore we acquire the camera object of the current frame, which allows for accessing its current pose. For each new frame, we check whether an augmented image was detected. If a new image was detected we add it to a list of currently tracked images. In the \textit{onDrawFrame} method we log the pose of the camera (of the current frame) and the poses of each detected augmented image together with a current timestamp of the smartphone.
Simultaneously, we log all the data from the smartphone's IMU including \textit{TYPE\_ACCELEROMETER}, \textit{TYPE\_GYROSCOPE}, \textit{TYPE\_MAGNETIC\_FIELD} and \textit{TYPE\_ROTATION\_VECTOR}.
Furthermore, we continuously request WLAN network scans via the \textit{WifiManager}.
For each session that we run, we create four separate logging files:
\begin{itemize}
    \item CameraPose.csv: holds the camera poses of each recorded frame
    \item LandmarkPose.csv: contains the current landmark poses (detected via augmented images) of each recorded frame
    \item Sensors.csv: contains all recorded sensors
    \item WiFi.csv: contains the results of the WLAN scans
\end{itemize}
Each entry in the files has an attached timestamp (in the same reference time) such that the entries can be matched. Note that the timestamps in CameraPose.csv \& LandmarkPose.csv are identical, while the other two files have to be aligned to the recorded frame time stamps.

During post-processing we compute a mapping to transform the camera poses from ACS to BCS by utilizing the continuously logged landmark positions. Subsequently, we assign a position to each entry of the Sensors.csv and WiFi.csv files by applying a time-based matching strategy.

\section{Mapping to building coordinate system (BCS)}\label{sec:postprocessing}

\subsection{Global mapping via least squares}

We identify a mapping from the positions of the landmarks in ACS to their corresponding positions in the BCS. This mapping can subsequently be used to transform all points of the trajectory.
The ACS has 3 dimensions, however, we discard the z-axis and assume that a single trajectory does not cover multiple floors. Further, the landmark height and the smartphone height are kept mostly constant.
The positions of our landmarks are continuously logged via the smartphone application (once they have been detected).
We compute the median position among all logged values for each landmark. Finally, we estimate the parameters of a similarity transformation via least squares minimization given the two point patterns as introduced in \cite{Umeyama91}.
In case the ACS stays mostly constant over the entire trajectory, this approach is suitable to achieve a highly accurate mapping.

\subsection{Locally optimal mapping of sub-trajectories}\label{sec:local_mapping}
ARCore continuously updates its internal world view, which results in an updated ACS. If these updates are significant, the logged positions of the static landmarks will deviate largely causing the global mapping strategy to fail.
In order to account for these updates, we compute locally optimal mappings, which are based on the most recent understanding (ACS) of ARCore.
We propose to partition the trajectory into sub-trajectories, which are
defined as the paths between each distinct landmark pair and compute individual transformations for each sub-trajectory based on the landmark positions in the most recent ACS.

\begin{figure}[ht]
    \centering
    \includegraphics[width=\linewidth]{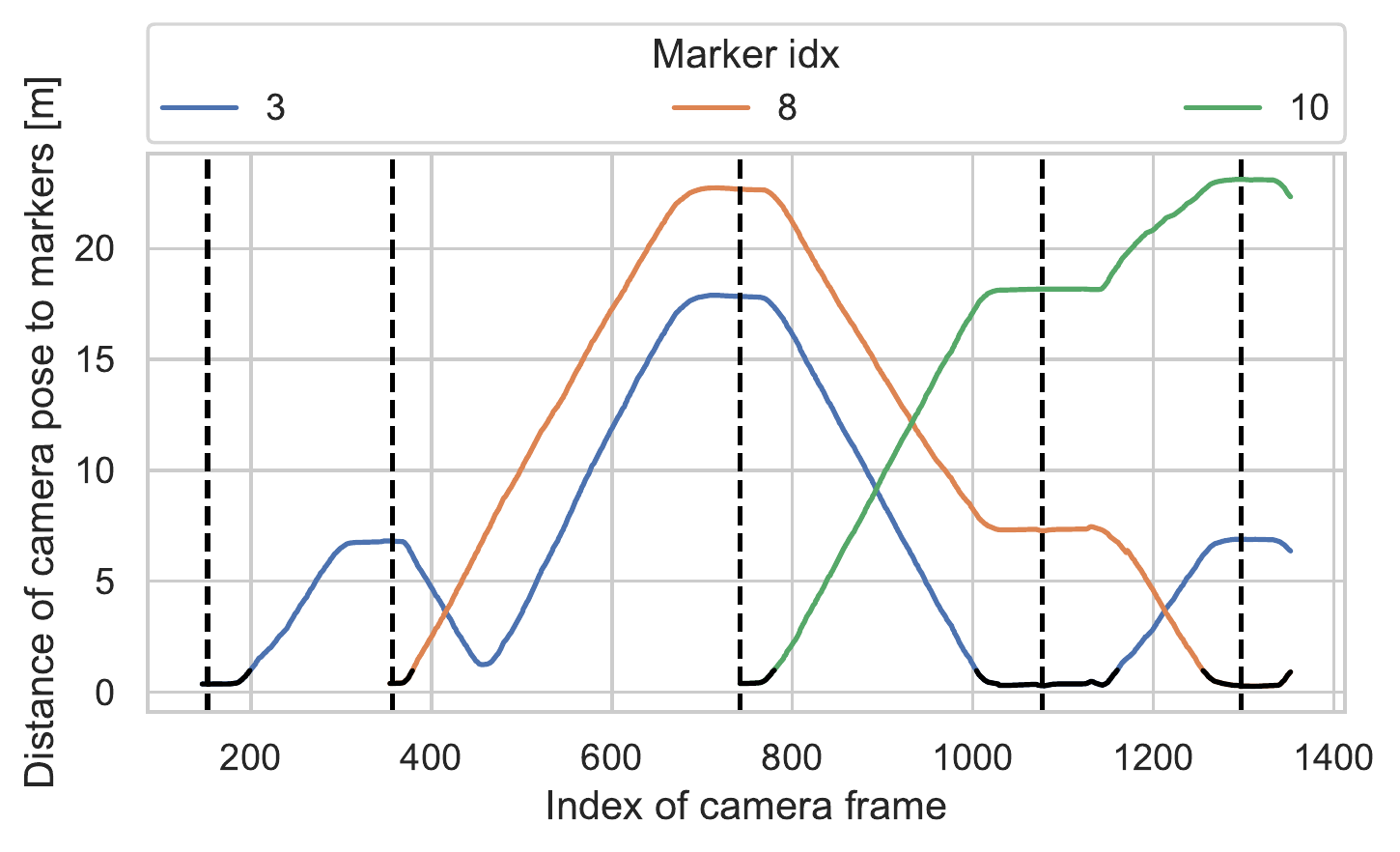}
    \caption{Detection of sub-trajectories via analyzing distance of camera to recorded reference markers.}
    \label{fig:subtrajectory_detection}
\end{figure}

\begin{figure}[ht]
    \centering
    \includegraphics[width=\linewidth]{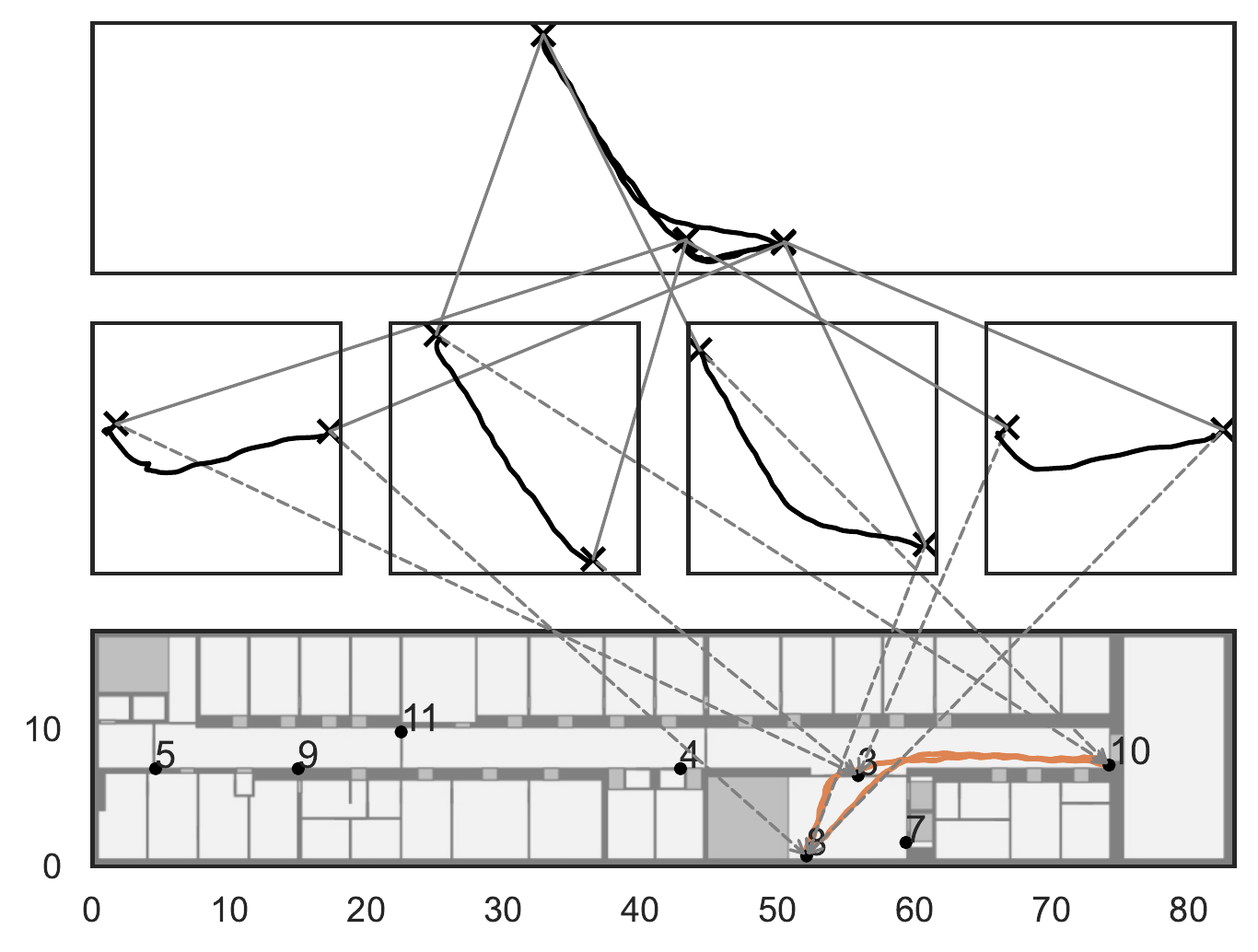}
    \caption{Local transformations of identified sub-trajecties between visited reference markers (black dots).}
    \label{fig:local_mapping}
\end{figure}

To identify the sub-trajectories, we determine the poses where to split the trajectory.
We detect possible candidates whose distance to a landmark lies below a predefined threshold (e.g. 1m). For each of these candidate intervals we compute the local minimum. Those minima identify the poses at which we will divide the trajectory. The process is illustrated in Figure \ref{fig:subtrajectory_detection}, the dotted lines represent the minima. Figure \ref{fig:local_mapping} shows how each sub-trajectory is locally mapped to the BCS.
The individual similarity transformations of the sub-trajectories are obtained by mapping the line between the landmark pair of the sub-trajectory from ACS to BCS by translation, uniform scaling and rotation. The resulting transformation is then applied to all poses of the given sub-trajectory.
Since this approach uses the most recent understanding of the world by utilizing the current positions of the landmarks instead of their global averages, this approach is much more robust against non-static ACS, as we will demonstrate in the evaluation.

\begin{figure}[ht]
    \centering
    \includegraphics[width=\linewidth]{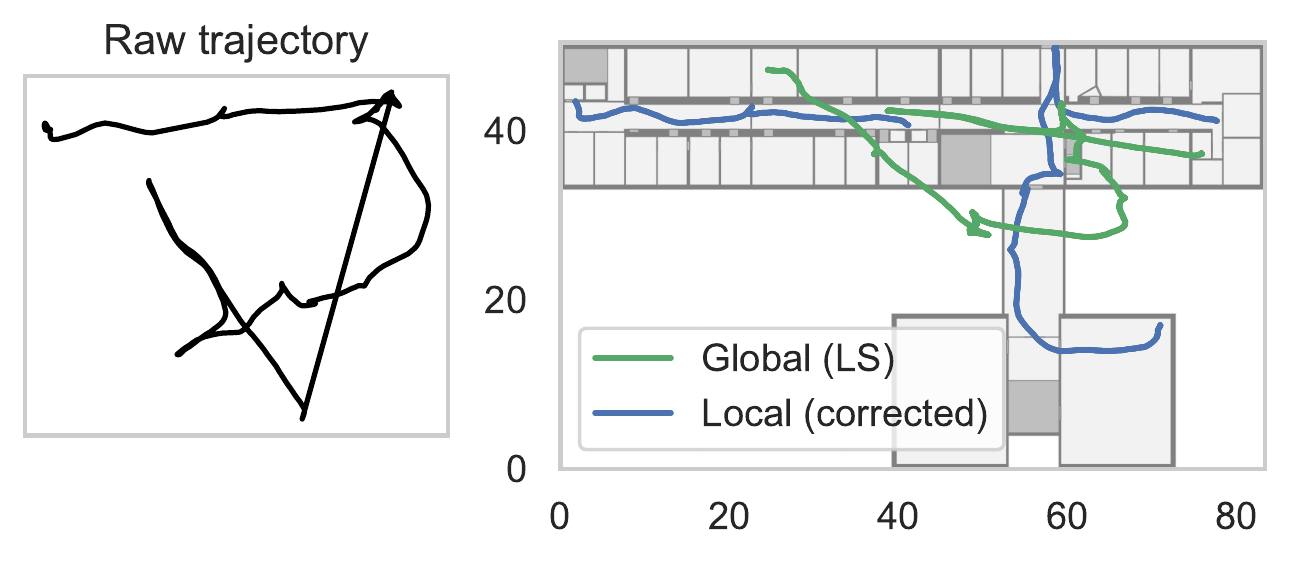}
    \caption{Flawed raw data cannot be mapped globally, whereas the local mapping still works to recover unimpaired parts of the data.}
    \label{fig:flailed_mapping}
\end{figure}

Since each sub-trajectory can be mapped individually, we can additionally discard certain segments that have a high probability of failure. We identify two main causes for failures: 1) ARCore loses track of previously visited landmarks. In particular, the camera pose and the landmark will be identical for a certain segment of frames. If we detect such cases, we set the recorded landmark position to "unknown". Given a sub-trajectory where ARCore loses track of all landmarks, we will discard the whole sub-trajectory; 2) It might happen that the camera pose jumps between two consecutive poses. These jumps are of a magnitude which is impossible to cover during a single frame. If jumps occur during a sub-trajectory, we will discard the whole sub-trajectory.
These two strategies allow us to identify parts of our walk where the VI-SLAM algorithm fails, while still recovering those parts where the algorithm was functioning as expected.
In contrast, when applying the global mapping, we would have to discard the entire trajectory. Small flawed parts of the trajectory distort the global mapping excessively, rendering the whole trajectory unusable. The qualitative mapping results of both strategies on a trajectory that contains severe ACS updates are exemplified in Figure \ref{fig:flailed_mapping}.

\section{Evaluation of VI-SLAM2tag labeling accuracy}\label{sec:evaluation}

The accuracy of the inferred position tags is evaluated using two distinct approaches for obtaining ground truth locations, which are described in the following.

\subsection{Accuracy via ground truth control points}
We place fixed control points on the floor and accurately register their positions in the BCS via geodetic surveying. Every time the user walks over a control point, he or she presses a button within the smartphone application such that the current timestamp is logged. Afterwards, we compare the post-processed position at the logged timestamp with the closest known control point.
\begin{figure*}[ht]
  \centering
  \subfloat[]{
  \makebox[.48\linewidth]{\includegraphics[width=.47\linewidth]{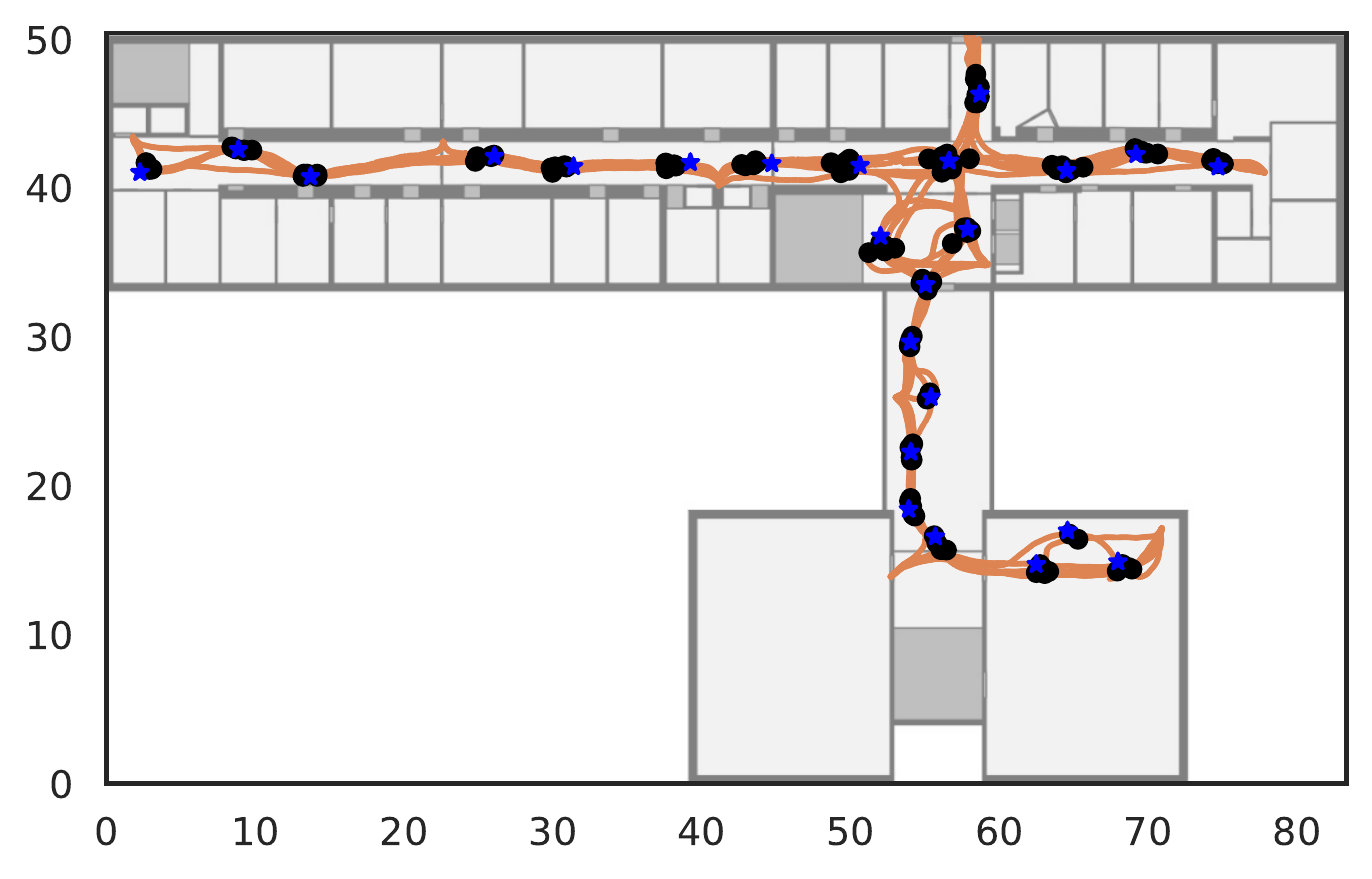}}
  }
  \subfloat[]{
  \makebox[.48\linewidth]{\includegraphics[width=.47\linewidth]{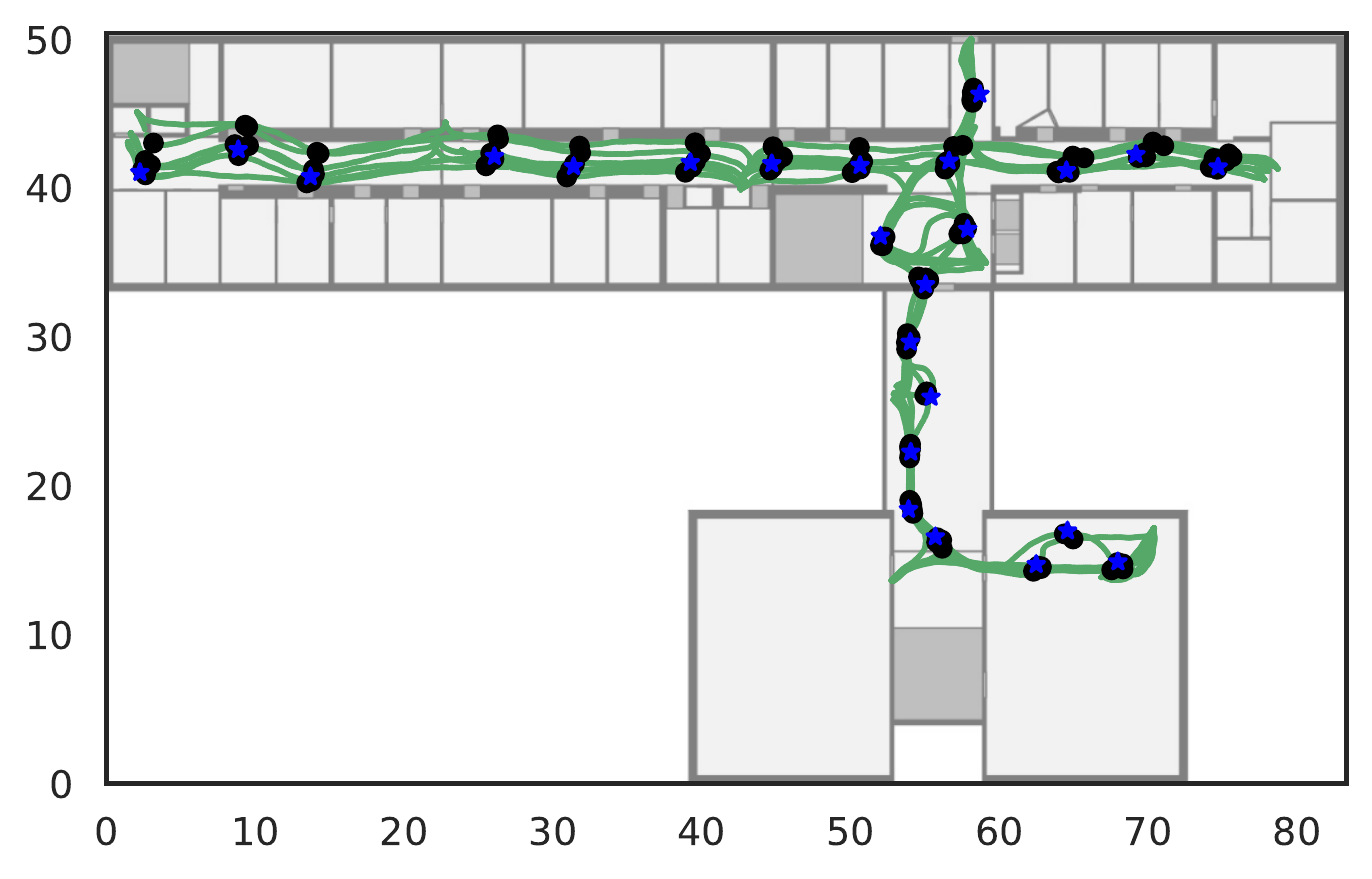}}
 }
  \caption{Visual comparison of mapped 15 minute trajectory. (a) shows local mapping strategy, (b) shows global mapping via least squares. The blue stars represent the control points and the black points show the computed position of the device at the timestamp of clicking the visited button. The numbers on x- and y-axis of the figure represent the local coordinates in meters.}\label{fig:mapping_cmp}
\end{figure*}
\begin{table}[h]
    \centering
    \begin{tabular}{llccc}
    \toprule
    \thead{Trajectory \\selection} & Metric &  \thead{Error of \\Local} &  \thead{Error of Local \\(corrected)} &  \thead{Error of \\Global (LS)} \\
    \midrule
    \multirow{3}{5em}{All trajectories} & Mean  &         0.91 &              0.53 &          2.80 \\
    & Median   &         0.63 &              0.49 &          0.54 \\
    & Maximum   &         4.51 &              0.87 &         20.78 \\
    \midrule
    \multirow{3}{5em}{Non-faulty trajectories} & Mean  &         0.51 &          - & 0.43 \\
    & Median   &         0.49 &    - &      0.40 \\
    & Maximum   &         0.72 &    - &      0.72 \\
    \bottomrule
    \end{tabular}
    \caption{Comparison of location error [m] of all mapping algorithms.}
    \label{tab:ref_pos_acc}
\end{table}
This evaluation strategy is meant for assessing the accuracy of the entire approach (including the post-processing) and roughly provides the final accuracy that is to be expected by the system.

\begin{figure}[h]
    \centering
    \includegraphics[width=\linewidth]{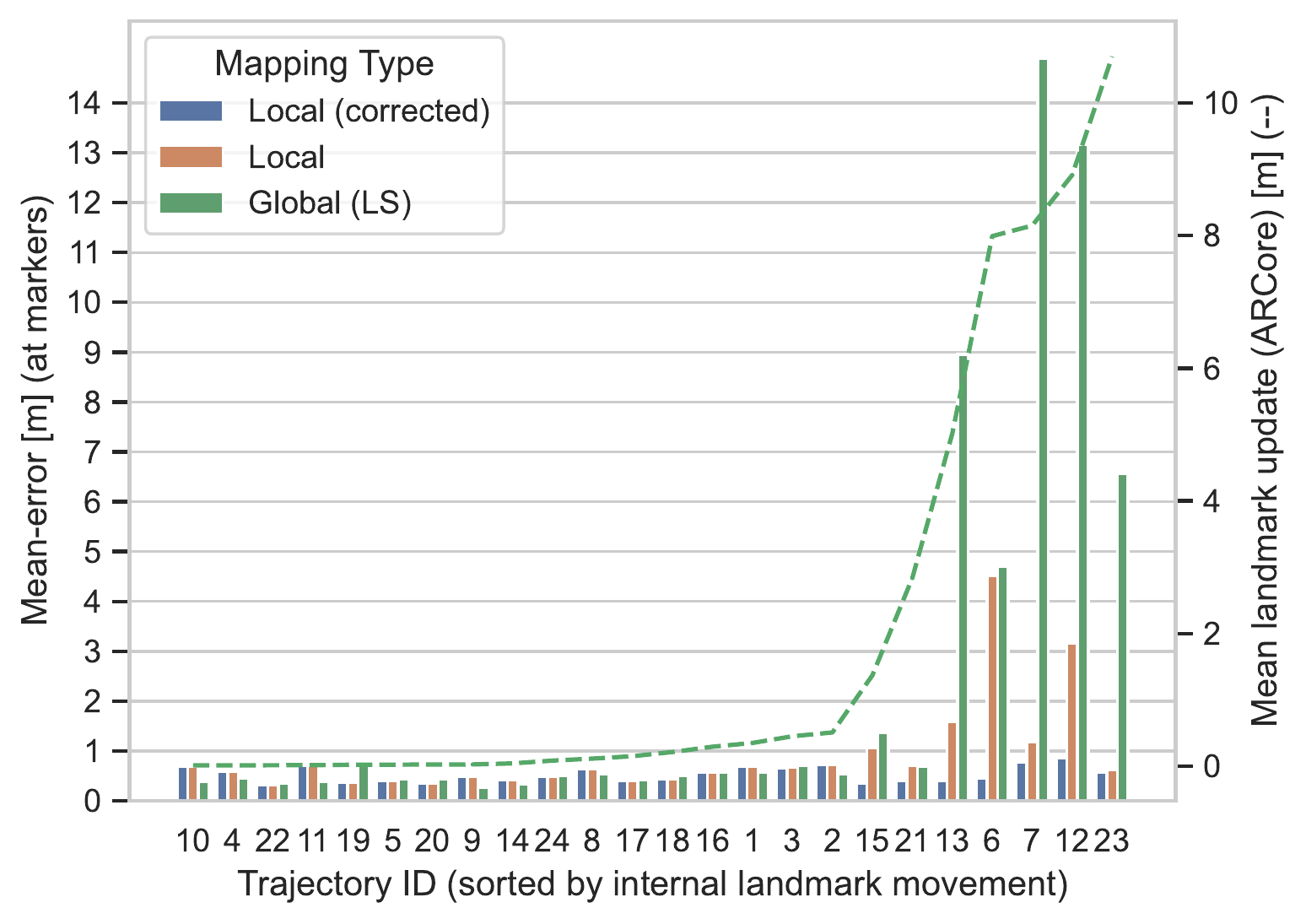}
    \caption{Relation of mean error and internal landmark update of ARCore.}
    \label{fig:error_update}
\end{figure}

In total we collected 24 reference trajectories on the 1st and 4th floor of our test building. We used two smartphones (LG V30 and OnePlus 6) and walked for a median duration of roughly 3 minutes per trajectory.
Table \ref{tab:ref_pos_acc} shows the obtained labeling accuracy, grouped into all trajectories and those where no localization error was detected by our mapping algorithm. The global mapping strategy has the highest accuracy on the non-faulty trajectories, since its least-squares mapping is more precise than locally mapping the sub-trajectories with only two corresponding points. Both mapping strategies achieve a median error of below 50 cm. However, on the flawed trajectories (e.g. updated ACS), the mean error of the global mapping rises to above 2m. In contrast, with the local mapping strategy, we can maintain a mean labeling error of just above 50 cm. We are interested in when the global mapping strategy fails, which is analyzed in Figure \ref{fig:error_update}.
It can be seen that the global mapping fails, if significant position updates of the logged landmark positions (fixed in real-world) happen due to ARCore updating its ACS. Since we utilize the median location of the landmarks for computing the global mapping, the high deviation in the logged landmark positions causes severe mapping inaccuracies and might even result in total failure as already shown in Figure \ref{fig:flailed_mapping}. In contrast, the local strategy utilizes the most recent location of the landmarks in the current ACS to obtain a mapping for each sub-trajectory, such that even for critical trajectories a high accuracy can be maintained.

To check whether the system is robust against drift effects, we walked a long trajectory of 15 minutes. The smallest error of 57 cm was achieved by the global mapping, which is comparable to those of the significantly shorter trajectories.
The obtained mapped trajectories are visualized in Figure \ref{fig:mapping_cmp} where the blue stars represent the positions of the control points and the black points represent the positions when the collector pressed the button to indicate that he or she just walked over a control point.

\subsection{Accuracy estimation via total station (TS) tracking}
The evaluation via comparison at ground truth control points can be used for roughly estimating the real-error among the whole trajectory.
\begin{figure}[h]
    \centering
    \includegraphics[width=\linewidth]{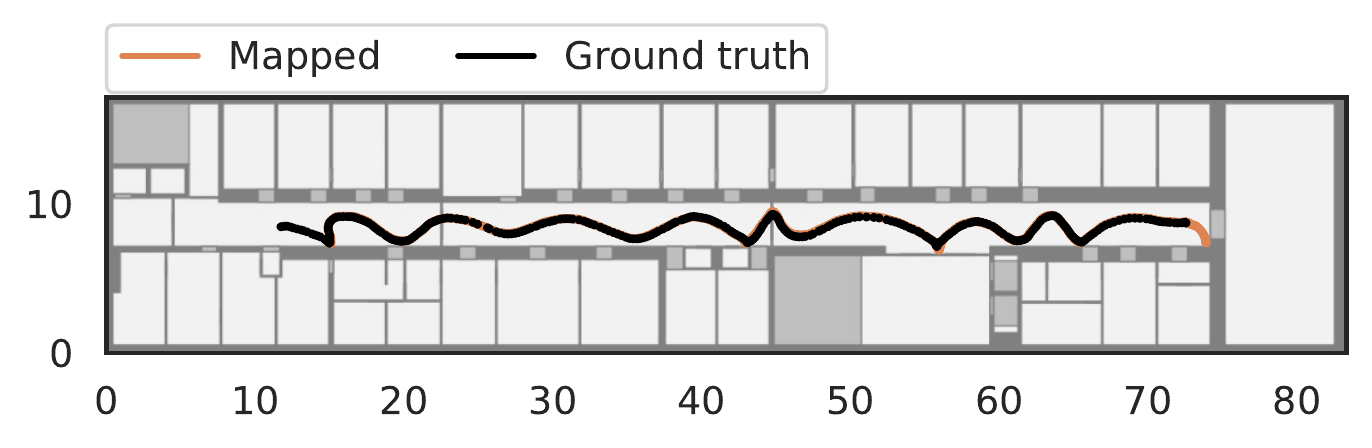}
    \caption{Overlay of computed trajectory and ground truth (via total station). Local coordinates are depicted in meters.}
    \label{fig:tachy_overlay}
\end{figure}
However, the measuring system itself, introduces a certain error, because manually clicking the button while walking over a control point has an inherent uncertainty. Furthermore, we only know the ground truth position when walking over the control point, which is only a snapshot given the whole trajectory. The system is mostly meant for WLAN fingerprint collection, therefore we assume that an accuracy below a meter is sufficient for automatic data labeling. However, we are also interested in assessing the accuracy of shorter trajectories with a more accurate ground truth measuring system that also allows for more frequent ground truth values.
Therefore, we utilize a total station (TS) (Leica Nova MS50) as an additional tracking system that provides accurate position information at a high sampling rate. The TS requires direct line-of-sight to the prisma installed at the monitored device (smartphone), which is why we will only apply it to trajectories in a corridor as depicted in Figure \ref{fig:tachy_overlay}.
The labeling accuracy of all collected trajectories is listed in Table \ref{tab:tachy_acc}. We collected several trajectories with changing velocities, which did not cause notable positioning errors, however, when instantly running, the algorithm fails, which results in sudden position jumps in the recorded trajectories. Those are detected and discarded by the local (corrected) mapping strategy, which is why its mean and max errors are significantly lower. The high labeling accuracy is qualitatively shown in Figure \ref{fig:tachy_overlay}, where the ground truth position (black) can hardly be distinguished from the post-processed trajectory (orange). During data collection, few people were present in the building, however, occasional acquaintances did not cause degrading performance. Out of curiosity, we simulated noise by a second persons that was heavily moving in the camera's view, which resulted in failure of the VI-SLAM algorithm. Therefore, it is advisable to not perform data collection during peak times of building occupancy.

\begin{table}[tbp]
    \centering
    \begin{tabular}{lccc}
    \toprule
    Metric &  \thead{Error of \\Local} &  \thead{Error of Local \\ (corrected)} &  \thead{Error of \\Global (LS)}\\
    \midrule
    Mean  &         0.73 &                   0.18 & -\\
    Median   &         0.23 &                   0.18 & -\\
    Maximum   &         2.68 &                   0.31 & -\\
    \bottomrule
    \end{tabular}
    \caption{Mapping error [m] for total station ground truth evaluation.}
    \label{tab:tachy_acc}
\end{table}

\section{The giaIndoorLoc dataset}\label{sec:dataset}

\begin{figure}[ht]
  \centering
  \subfloat[Fingerprints per device]{
  \makebox[.47\linewidth]{\includegraphics[trim=0 20 0 19, clip, height=3.9cm]{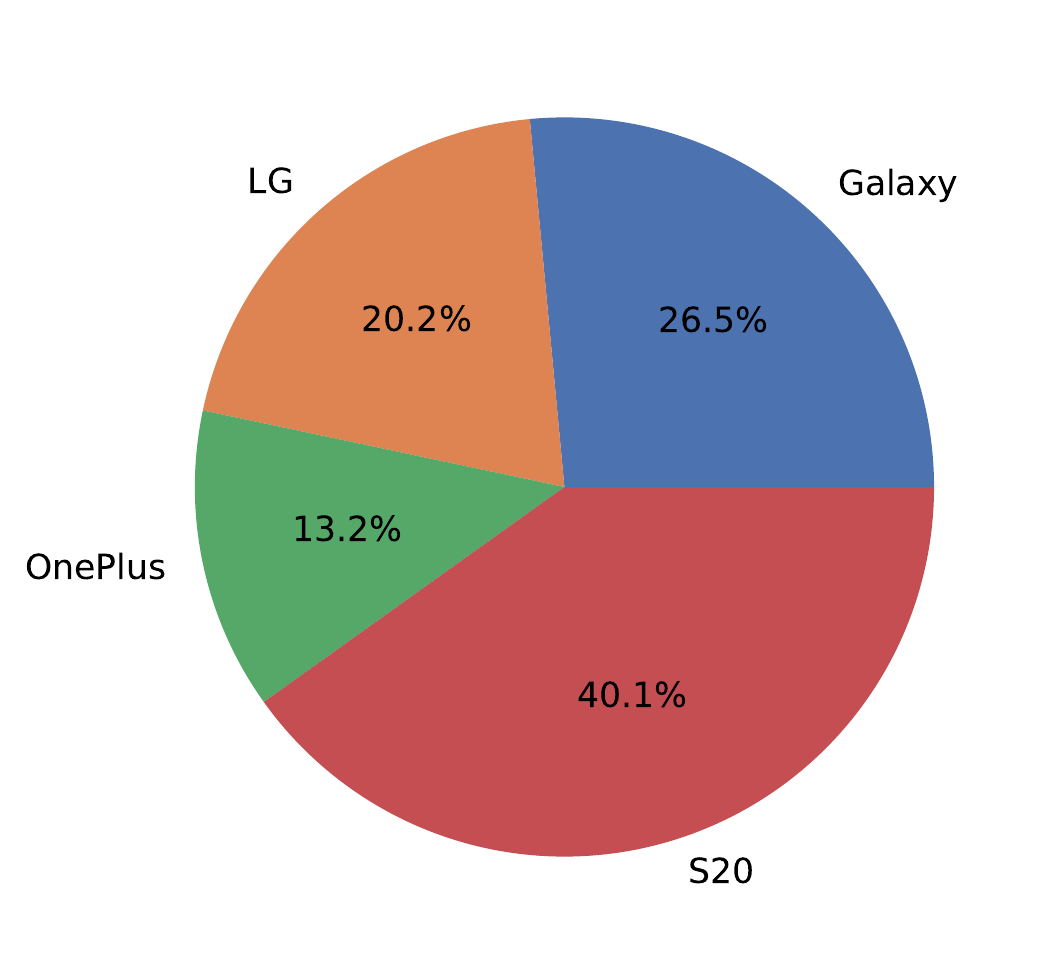}}
  }
  \subfloat[Fingerprints per floor]{
  \makebox[.47\linewidth]{\includegraphics[trim=0 20 0 19, clip, height=3.9cm]{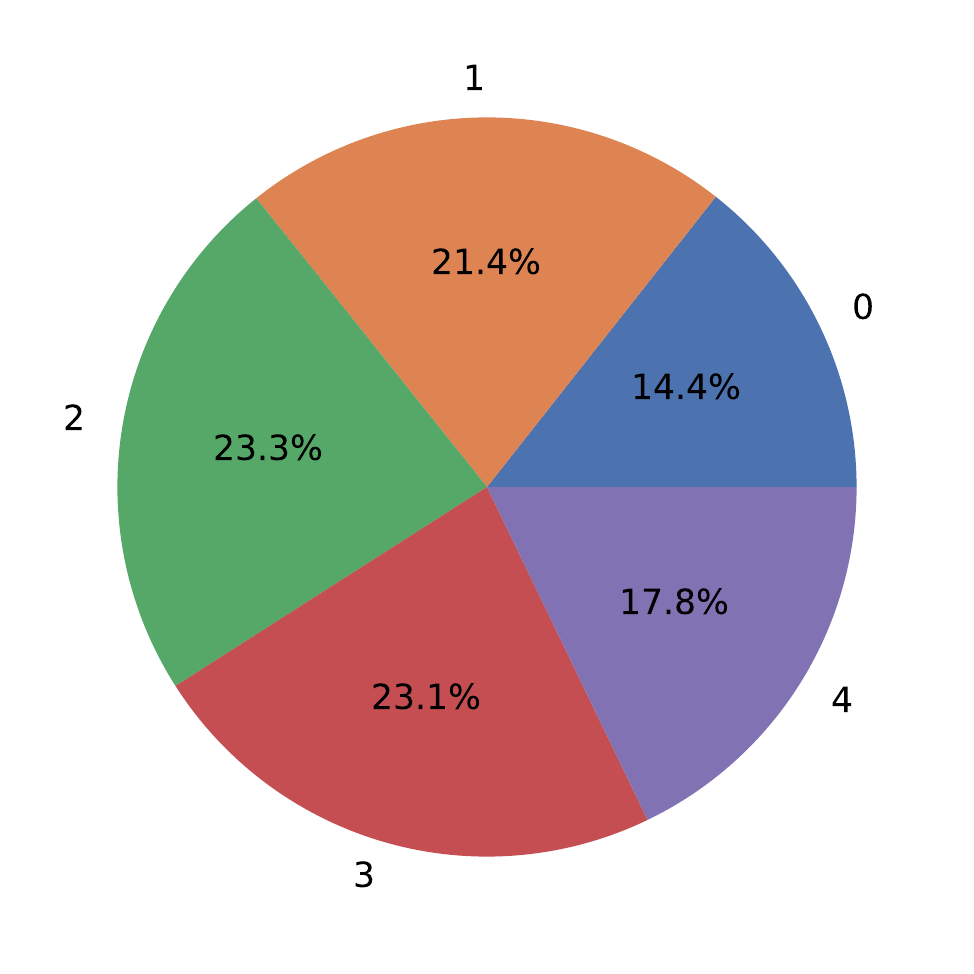}}
 }
  \caption{Distribution of fingerprints across devices and floors of the building.}\label{fig:dataset_stats}
\end{figure}

\begin{figure}[h]
    \centering
    \includegraphics[width=\linewidth]{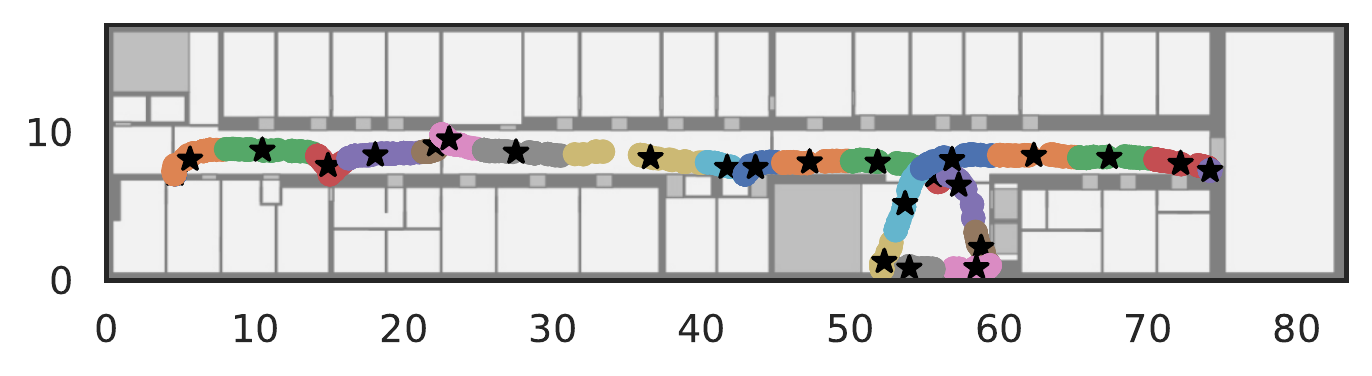}
    \caption{Visualization of obtained WLAN fingerprints of example trajectory. Each dot represents the smartphone position at the time of an obtained AP RSS reading. A single color is chosen per network scan. The black stars represent the average position of each scan, which is used for annotation.}
    \label{fig:dynamic_fp_labels}
\end{figure}

With the help of the introduced VI-SLAM2tag system we collect a dataset at our university building that spans 5 floors. The entire data collection took less than 2 hours for 2 persons, which includes the setup time at each floor (placing landmark images). The site survey resulted in a total of 2049 auto-labeled fingerprints (after automatically discarding possibly wrong labels), which stemmed from 4 different smartphones (OnePlus 6, LG V30, Samsung S20 Ultra, Samsung Galaxy A52S). Statistics on the dataset are depicted in Table \ref{tab:dataset_stats} and Figure \ref{fig:dataset_stats}.
\begin{table}[ht]
    \centering
    \begin{tabular}{lcccc}
    \toprule
    Device & \thead{Number of \\trajectories} & \thead{Duration \\ \text{[min]}} & \thead{Walked \\distance [m]} & \thead{Average \\velocity [m/s]} \\
    \midrule
    Galaxy  &                     24 &          54.77 &             2559.39 &                   0.99 \\
    LG      &                     18 &          77.56 &             3701.38 &                   0.92 \\
    OnePlus &                     24 &          51.49 &             2351.40 &                   0.94 \\
    S20     &                     15 &          68.18 &             2624.85 &                   0.98 \\
    \bottomrule
    \end{tabular}
    \caption{Dataset statistics.}
    \label{tab:dataset_stats}
\end{table}
The dataset allows for training supervised models at several tasks including WLAN- or magnetic-based fingerprinting for absolute position estimation but also relative position estimation via end-to-end deep learning on IMU data as proposed in \cite{CLW+21}. In the following we will demonstrate how to the dataset can be used for training WLAN fingerprinting-based models. When requesting WLAN scans via the Android operating system, each recorded entry is tagged with the timestamp of collection. Those can slightly differ for the seen APs in a single scan as the scan might require several seconds depending on the hardware of the device. Since we obtain high frequent pseudo-ground truth labels via VI-SLAM2tag, we are technically able to assign accurate positions for each AP entry within a single network scan as depicted in Figure \ref{fig:dynamic_fp_labels}. In order to train our model in the standard way of having a single position label for each network scan (fingerprint), we average the recorded positions per scan to compute a global label for each scan (black stars). However, we want to emphasize that the dynamic collection in combination with the high frequent position annotations enable further advanced fingerprint generation following spatial data augmentation schemes.

\subsection{Baseline fingerprinting performance}
For assessing a reference performance on the dataset we choose two static models based on neural networks. A standard multi-layer perceptron with a 3D-regression output and the recently proposed multi-CEL model \cite{LB22a}, which was explicitly designed for multi-floor localization using a single neural network.
We utilize multiple train/test splits to answer two questions: i) what is the general localization accuracy across all devices by a random split into 30\% test data and ii) how well does the algorithm generalize across devices by using the data of a single phone for testing only.

\begin{table}[h]
    \setlength{\tabcolsep}{4pt}
    \centering
    \begin{tabular}{llccc}
    \toprule
    Test device & Model &  floor\_ACC &  Mean error [m] &  Median error [m] \\
    \midrule
    \multirow{2}{5em}{Random split (30\%)} & mCEL &      0.992 &        2.08 &          1.52 \\
    & 3Dreg &      0.988 &        2.59 &          2.12 \\
    \midrule
    \multirow{2}{3em}{Galaxy} & mCEL &      0.987 &        2.68 &          2.02 \\
    & 3Dreg &      0.979 &        3.36 &          2.53 \\
    \midrule
    \multirow{2}{2em}{LG} & mCEL &      0.992 &        2.59 &          1.71 \\
    & 3Dreg &      0.967 &        2.93 &          1.94 \\
    \midrule
    \multirow{2}{4em}{OnePlus} & mCEL &      0.968 &        3.06 &          1.76 \\
    & 3Dreg &      0.981 &        3.80 &          2.82 \\
    \midrule
    \multirow{2}{2em}{S20} & mCEL &      0.999 &        2.05 &          1.74 \\
    & 3Dreg &      0.993 &        2.84 &          2.37 \\
    \midrule
    \bottomrule
    \end{tabular}
    \caption{Performance on several test data splits.}
    \label{tab:perf_static_fp}
\end{table}

The results are shown in Table \ref{tab:perf_static_fp}. Irrespectively of the train/test split, m-CEL outperformed the 3D-regression model with respect to floor detection accuracy and mean/median positioning error. It achieves a mean positioning error of roughly 2m. Furthermore, the model is able to generalize well across different devices, reaching a similar mean positioning accuracy as compared to the random split.

\section{Conclusion}\label{sec:conclusion}
We presented VI-SLAM2tag for dynamically collecting auto-labeled fingerprints.
It consists of i) a smartphone application that uses ARCore for tracking the position during fingerprint collection and ii) a post-processing module that transforms the position tags to the target coordinate system while being robust against coordinate system updates of ARCore.
The obtained labeling accuracy was evaluated using two ground truth (GT) measuring systems. The control point-based GT system showed a labeling accuracy of roughly 50 cm even for trajectories of up to 15 minutes. To exclude measurement uncertainties, we additionally utilized a total station which indicates that for short trajectories a labeling accuracy of below 20 cm can be expected.
This is sufficient for WLAN fingerprinting, since the prediction accuracy lies well above the labeling accuracy.
Within 2 hours we collected a labeled dataset within a 5-floor building of our university and demonstrated that trained models achieved a positioning accuracy of roughly 2m. In future work we want to leverage the high frequent labels of the dynamically collected fingerprints via spatial data augmentation techniques.
VI-SLAM2tag is made openly accessible to simplify future deployments of fingerprinting-based indoor localization systems.

\bibliographystyle{IEEEtran}
\bibliography{IEEEabrv,ref}

\end{document}